\documentstyle[12pt]{article}
\begin{document}
\setlength{\oddsidemargin}{-0.1cm}
\setlength{\topmargin}{-1.cm}

\title{A critical assessment of the Self-Interaction Corrected
Local Density Functional method and its algorithmic implementation }
\author{S. Goedecker,
\\Max-Planck Institute for Solid State Research, Stuttgart, Germany
\\
\\C. J. Umrigar,
\\Cornell Theory Center and Laboratory of Atomic and Solid State Physics,
\\ Cornell University, Ithaca, NY 14853, USA }
\date{}
\maketitle

\Large ABSTRACT \normalsize \newline
We calculate the electronic structure of several atoms and small molecules by direct
minimization of the Self-Interaction Corrected Local Density Approximation (SIC-LDA) functional.
To do this we first derive an expression for the gradient of this functional under the
constraint that the orbitals  be orthogonal and show that previously given
expressions do not correctly incorporate this constraint. In our atomic calculations
the SIC-LDA yields total energies, ionization energies and charge densities that
are superior to results obtained with the
Local Density Approximation (LDA).  However, for molecules SIC-LDA gives
bond lengths and reaction energies that are inferior to those obtained from LDA.
The nonlocal BLYP functional, which we include as a representative GGA functional,
outperforms both LDA and SIC-LDA for all ground state properties we considered.

\pagebreak
\section*{Introduction}
The Local Density Approximation (LDA) \cite{dft} has become one of the most popular tools
for electronic structure calculations. The reason for this is that it gives good
accuracy for structural properties and is computationally less costly than
traditional quantum chemistry methods such as Hartree-Fock, Configuration Interaction
and Coupled Cluster methods. With the rapid increase in computer power and the
development of low complexity algorithms, the limits on the system size are
being pushed up steadily.  However, in many cases LDA is not sufficiently accurate.
A primary concern is therefore to improve upon the accuracy of
the LDA approximation at the expense of a moderate increase in computational cost.
The Generalized Gradient Approximations  such as the Becke-Lee-Yang-Parr (BLYP) scheme
\cite{gga,becke} fall into this category and are now widely used.
Numerous other schemes to improve upon the LDA scheme can been found in the literature~\cite{beyondLDA},
but very few of them have been systematically tested in atomic and molecular calculations.
A major deficiency of the LDA and also to a lesser extent of the GGA is the fact, that
there is an unphysical self-interaction in these functionals. To cure this, several years
ago Perdew and Zunger proposed a scheme \cite{pz:sic} where self-interaction
terms are subtracted out in a straightforward way (SIC-LDA). Even though this scheme
is appealing because of its conceptional simplicity, it has not been widely used
possibly because it leads to a numerically more complicated
scheme since the potential becomes orbital dependent. The minimization of this
functional therefore cannot anymore be considered as an eigenvalue problem in a
self-consistent potential and the total energy is not any more invariant under unitary
transformations among the occupied orbitals. In this paper, we first derive the gradient
of the SIC-LDA functional, which is necessary for minimization algorithms \cite{mimi,t:prec}, 
and then present results that we obtained for atomic and molecular systems.

Studies of periodic systems using SIC-LDA have mainly concentrated on systems 
where the LDA approximation gives qualitatively wrong results (such 
as transition metal oxides \cite{transox}) and 
where the electronic structure undergoes qualitative changes in response to 
changing external conditions \cite{svane,cerium}.  Most of these calculations involving 
heavy atoms were done with the LMTO method. The gap in insulators is 
also improved compared with the LDA case \cite{gap}.
In the case of atoms and molecules most papers \cite{peder,harri,white} 
investigated
the properties of excited states of atoms and molecules at the experimental
geometry and found that the SIC-LDA improves upon LDA.
Instead, we concentrate on the ground state properties such as the
equilibrium geometries of small molecules and find that SIC-LDA is less accurate
than LDA.

\section*{The SIC-LDA functional and its gradient}
In all the following formulae we consider the orbitals
$\Psi_i$ to be real and the subscripts of the orbitals run over all
the occupied orbitals.
The SIC-LDA functional is given by
\begin{eqnarray}
E_{\rm tot}[\Psi_i({\bf r})] & = & \sum_i \int \Psi_i({\bf r}) \left(-\frac{1}{2} \nabla^2 + V_{\rm ext} \right) \Psi_i({\bf r}) d{\bf r}\\
  & + & \frac{1}{2} \int  \int  \frac{ \rho({\bf r}) \rho({\bf r}' ) }{|{\bf r} - {\bf r}' |}  d{\bf r} d{\bf r}'
 - \frac{1}{2} \sum_i \int \int \frac{ \rho_i({\bf r} ) \rho_i({\bf r}' ) }{|{\bf r} -{\bf r}' |} d{\bf r} d{\bf r}'  \nonumber \\
   & + & \int \epsilon(\rho({\bf r} )) \rho({\bf r} ) d{\bf r}
   - \sum_i \int \epsilon(\rho_i({\bf r} )) \rho_i({\bf r} ) d{\bf r} \nonumber
\end{eqnarray}
where
$$ \rho_i({\bf r} ) = \Psi_i({\bf r} ) \Psi_i({\bf r} ), \;\;\;
\rho({\bf r} ) = \sum_i \rho_i({\bf r} ) $$
and $\int \epsilon(\rho) \rho d{\bf r} $ is the local approximation of the exchange correlation functional.

We want to obtain its gradient under the constraint that the orbitals be
orthonormal. Following the ideas outlined by Arias et al. \cite{a:min} ,
we consider a
more general functional that is also defined with respect to nonorthogonal
functions $\Psi_i$. First we construct a set of orthonormal orbitals $\tilde{\Psi}_i$
by a symmetric L\"{o}wdin
orthogonalization of the non-orthogonal set $\Psi_i$
\begin{equation}
 \tilde{\Psi}_i = \sum_j S_{i,j}^{-1/2} \Psi_{j}
\end{equation}
where $S_{i,j} = <\Psi_i|\Psi_j> = \int \Psi_i({\bf r} ) \Psi_j({\bf r} ) d{\bf r} $ is the overlap
matrix among the occupied orbitals.
The functional we are interested in is just the SIC-LDA functional evaluated for
the orthonormal orbitals $\tilde{\Psi}_i$.
In our actual calculation, for reasons of numerical stability, we use orthogonal orbitals.
Hence, in our derivations it is necessary to consider only infinitesimally
nonorthogonal orbitals . Then $S^{-1/2} = (I+(S-I))^{-1/2} \approx  I -(1/2)(S-I) $
and Eq.~(2) simplifies to
\begin{equation}
\tilde{\Psi}_i = \sum_j (\frac{3}{2} \delta_{i,j} - \frac{1}{2} S_{i,j}) \Psi_j
\end{equation}
The gradient of the total functional is then obtained by applying the chain rule:
\begin{equation}
\frac{\partial E}{ \partial \Psi_i({\bf r} )} =
        \sum_j \int \frac{\partial E}{ \partial \tilde{\Psi}_j({\bf r}' )}
                    \frac{\partial \tilde{\Psi}_j({\bf r}' ) }{ \partial \Psi_i({\bf r} )} d{\bf r}'
\end{equation}
where each part of Eq.~(4) can easily be calculated. Denoting the unconstrained gradient by
$d_j({\bf r} )$ we obtain:
\begin{eqnarray}
d_j({\bf r} )  =  \frac{1}{2} \frac{\partial E}{ \partial \tilde{\Psi}_j({\bf r} )} & =  &
     \left( -\frac{1}{2} \nabla^2 + V_{\rm ext} \right) \tilde{\Psi}_j({\bf r} )  \\
   & + &   \left( \int  \frac{\rho({\bf r}' ) }{|{\bf r} -{\bf r}' |} d{\bf r}'
 -  \int  \frac{\rho_j({\bf r}' ) }{|{\bf r} -{\bf r}' |} d{\bf r}' \right) \tilde{\Psi}_j({\bf r} ) \nonumber \\
   & + & \left( \mu(\rho({\bf r} ))  - \mu(\rho_j({\bf r} )) \right) \tilde{\Psi}_j({\bf r} ) \nonumber \\
   & = & H_j \tilde{\Psi}_j({\bf r} ) \; , \nonumber
\end{eqnarray}
where the orbital dependent Hamiltonian $H_j$ is
\begin{eqnarray}
H_j & = & (-\frac{1}{2} \nabla^2 + V_{\rm ext}) \nonumber  \\
   & + &   \int  \frac{ \rho({\bf r}' ) }{|{\bf r} -{\bf r}' |} d{\bf r}'
   -   \int  \frac{  \rho_j({\bf r}' ) }{|{\bf r} -{\bf r}' |} d{\bf r}' \nonumber  \\
   & + & \mu(\rho({\bf r} ))  - \mu(\rho_j({\bf r} )). \nonumber
\end{eqnarray}
The second part  of Eq.~(4) gives
\begin{eqnarray}
\frac{\partial \tilde{ \Psi}_j({\bf r}' ) }{ \partial \Psi_i({\bf r} )} & = &
\frac{3}{2} \delta_{i,j} \delta({\bf r} -{\bf r}' ) -\frac{1}{2} S_{i,j} \delta({\bf r} -{\bf r}' )
- \frac{1}{2} \sum_l
\frac{\partial}{\partial \Psi_i({\bf r} )} \int \Psi_j({\bf r}'' ) \Psi_l({\bf r}'' ) d{\bf r}''  \Psi_l({\bf r}' ) \nonumber \\
 & = & \frac{3}{2} \delta_{i,j} \delta({\bf r} -{\bf r}' ) -\frac{1}{2} S_{i,j} \delta({\bf r} -{\bf r}' )
- \frac{1}{2} \sum_l \left
(\delta_{i,j} \Psi_l({\bf r} )+ \delta_{i,l} \Psi_j({\bf r} ) \right)  \Psi_l({\bf r}' ) \nonumber \\
& = & \delta_{i,j} \delta({\bf r} -{\bf r}' )
- \frac{1}{2} \Psi_i({\bf r}' )  \Psi_j({\bf r} )
- \frac{1}{2} \delta_{i,j} \sum_l \Psi_l({\bf r} )  \Psi_l({\bf r}' ) \; .
\end{eqnarray}
In the last transformation step, we have used the fact that we calculate the
derivative for a set of orthonormal orbitals  and therefore $S=I$.
In order to take account of the orthogonality constraint, we are of course
allowed to put $S=I$ only after calculating the derivative.
Finally we obtain the gradient expression
\begin{eqnarray}
\frac{1}{2} \frac{\partial E}{ \partial \Psi_i({\bf r} )} = d_i({\bf r} )
  & - & \frac{1}{2} \sum_j \left(\int d_j({\bf r}' ) \Psi_i({\bf r}' ) d{\bf r}' \right) \Psi_j({\bf r} ) \nonumber \\
  & - & \frac{1}{2} \sum_j \left(\int d_i({\bf r}' ) \Psi_j({\bf r}' ) d{\bf r}' \right) \Psi_j({\bf r} ) .
\label{gradient}
\end{eqnarray}
The above gradient expression is different from what is found in the
literature \cite{harri,fois,svane}.
\begin{equation}
\frac{1}{2} \frac{\partial E}{ \partial \Psi_i({\bf r} )} = d_i({\bf r} )
  - \sum_j \left(\int d_i({\bf r}' ) \Psi_j({\bf r}' ) d{\bf r}' \right) \Psi_j({\bf r} )
\label{gradient_wrong}
\end{equation}
In numerical applications, it was apparent that the gradient expression
in Eq.~\ref{gradient_wrong} does
not lead to the correct minimum \cite{fois} of the functional. However, nobody seems
to have drawn the logical conclusion that Eq.~~\ref{gradient_wrong} is not the
correct gradient of the LDA-SIC functional.
Instead people added a second minimization step which is based on
a relation derived by Pederson et al \cite{peder}
\begin{equation}
< \Psi_i| H_j \Psi_j> = <H_i \Psi_i | \Psi_j> .
\label{pederson}
\end{equation}
This relation follows immediately by considering infinitesimal unitary transformations
among the occupied orbitals and putting the gradient of the SIC-LDA functional
with respect to these transformations equal to zero.
We see that Eqs.~\ref{gradient_wrong} and \ref{pederson}
together are equivalent to Eq.~\ref{gradient}.
These difficulties appear only in the SIC-LDA case; in the LDA case the
potential is not orbital dependent  and therefore the two gradient expressions
are identical.

The results of the following two sections were obtained by direct minimization of
the SIC-LDA functional, using the correct gradient~\ref{gradient}.
The DIIS method \cite{h:diis} was used
as a convergence accelerator. In the molecular case the gradient was preconditioned
using the scheme by Teter et al. \cite{t:prec}, in the atomic case the
operator $Im(\frac{1}{H-z})$ was used as a preconditioner \cite{g:low} ,
where z is a suitably chosen complex energy. All the calculations were done
for a spin unpolarized system where the spin up orbital
is required to have the same spatial form as the matching spin down electron orbital,
$\Psi_{2i-1}({\bf r} )=\Psi_{2i}({\bf r} )$

To calculate the equilibrium geometries of small molecules, we relaxed the
atoms in the direction of the forces until the forces vanished. The forces
in the SIC-LDA scheme are given by the Hellman-Feynman theorem. This might
not be quite obvious, since in the usual derivation of the Hellman-Feynman theorem
one takes advantage of the fact, that the orbitals are eigenfunctions of the
self-consistent Hamiltonian, which is not the case in the SIC-LDA scheme.
We therefore give in the following a derivation of the Hellman-Feynman theorem
which does not require the orbitals to be eigenfunctions, but uses only the
fact that the orbitals minimize the total energy for some fixed positions of the
nuclei.

Let us consider a set of orbitals  $\Psi_i({\bf r} ,\vec{{\bf R}}')$ which are the solutions
of the SIC-LDA equations for a molecule, whose atomic positions are given
by $\vec{{\bf R}}'$. We consider $\vec{{\bf R}}'$ to be a super vector, i.e. for a system
comprising $N$ atoms it will have $3N$ components. By construction these
orbitals  are orthonormal for all values of $\vec{{\bf R}}'$. Let us now
consider the SIC-LDA total energy $E_{\rm tot}[\Psi_i({\bf r} ,\vec{{\bf R}}');\vec{{\bf R}}]$
for a set of atomic positions $\vec{{\bf R}}$. The dependence on $\vec{{\bf R}}$ stems
from the fact that the external potential $V_{\rm ext}$ depends on the
atomic positions $\vec{{\bf R}}$. Obviously the functional will
be minimized if $\Psi_i({\bf r} ,\vec{{\bf R}}')=\Psi_i({\bf r} ,\vec{{\bf R}})$ and its gradient
with respect to $\vec{{\bf R}}'$ will vanish at that point.
\begin{equation}
\left. \frac{\partial E_{\rm tot}[\Psi_i({\bf r} ,\vec{{\bf R}}');\vec{{\bf R}}]}
            {\partial \vec{{\bf R}}'} \right|_{\vec{{\bf R}}'=\vec{{\bf R}}} = 0
\end{equation}
Let us now calculate the force acting on the nuclei which is given by
\begin{equation}
\frac{\partial E_{\rm tot}[\Psi_i({\bf r} ,\vec{{\bf R}});\vec{{\bf R}}]}{\partial \vec{{\bf R}}}  =
\left. \frac{\partial E_{\rm tot}[\Psi_i({\bf r} ,\vec{{\bf R}});\vec{{\bf R}}'']}
            {\partial \vec{{\bf R}}''} \right|_{\vec{{\bf R}}''=\vec{{\bf R}}}   +
\left. \frac{\partial E_{\rm tot}[\Psi_i({\bf r} ,\vec{{\bf R}}');\vec{{\bf R}}]}
            {\partial \vec{{\bf R}}'} \right|_{\vec{{\bf R}}'=\vec{{\bf R}}}
\end{equation}
The first term on the right side of Eq.~(11)
takes into account that the external potential $V_{\rm ext}$ depends on the atomic
positions $\vec{{\bf R}}$ but freezes the dependence of the orbitals  on $\vec{{\bf R}}$.
It is given by
\begin{equation}
\sum_i \int
\Psi_i({\bf r} ,\vec{{\bf R}}) \frac{\partial V_{\rm ext}}{\partial \vec{{\bf R}}} \Psi_i({\bf r} ,\vec{{\bf R}}) d{\bf r}
\end{equation}
The second term freezes the dependence on the external potential but takes into
account the dependence of the orbitals . This second term is however
equal to the left hand side of Eq.~(10) and therefore vanishes. So we are just left with
the first term which can be immediately transformed to the usual
Hellman-Feynman form.

\section*{Atomic results}
In their original paper, Perdew and Zunger \cite{pz:sic} do several atomic
calculations. They however did not minimize the SIC-LDA functional under the
orthogonality constraint, but instead employed the eigen orbitals of the
orbital dependent potential.  Since the orbitals are the eigenfunctions
for different Hamiltonians, they are not orthogonal but they assume that
this will not change their results appreciably.
We therefore repeated some of the atomic calculations to check this assumption
and in fact find it to be true.


The orbitals that one obtains from a minimization solution differ in two
important aspects from the atomic orbitals  obtained in a non-orbital
dependent potential. First their nodal structure is different as shown in Fig.1.
All orbitals  (even the 1s core states) have the same number of nodes.
Second all the orbitals  decay with the same exponent (see Fig.~2).
A set of orbitals  with a behaviour appropriate for a local potential can
however be obtained by forming the new linear combinations
\begin{equation}
\Phi_i = \sum_j C_{ij} \Psi_j
\label{eigen}
\end{equation}
where $C$ is the matrix containing the eigenvectors of the (hermitian) matrix
$<\Psi_i | H_j \Psi_j>$.

    \begin{figure}             
     \begin{center}
      \setlength{\unitlength}{1cm}
       \begin{picture}( 8.,7.0)           
        \put(-1.,0.){\includegraphics{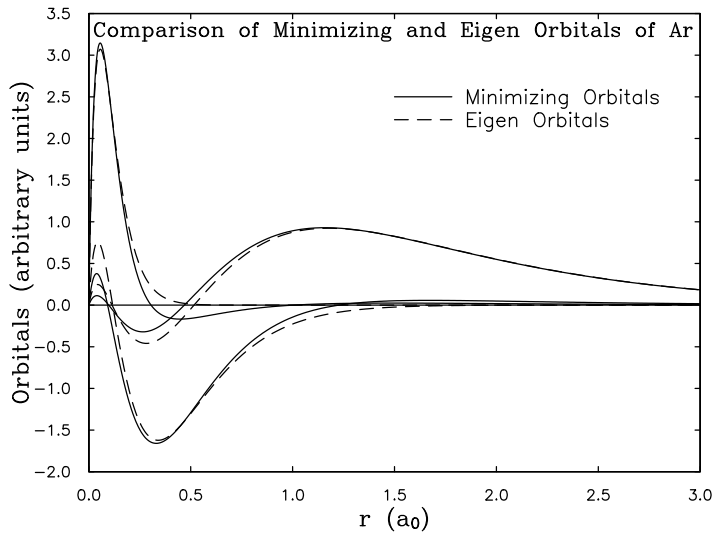}}   
       \end{picture}
       \caption{ 1s, 2s, and 3s orbitals  of Argon. Solid line:
                  The orbitals  $\Psi$ that minimize the SIC-LDA
                 functional, dashed line: the linear combinations $\Phi$ of Eq.13.  }
      \end{center}
      \label{psi}
     \end{figure}

    \begin{figure}             
     \begin{center}
      \setlength{\unitlength}{1cm}
       \begin{picture}( 8.,7.0)           
        \put(-1.,0.){\includegraphics{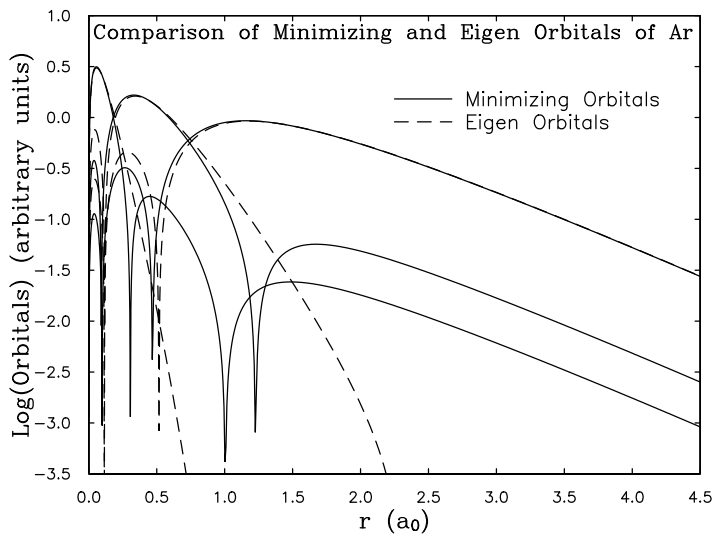}}   
       \end{picture}
       \caption{ Same as Fig.1 but on a logarithmic scale.
       The 3s orbitals $\Psi$ and $\Phi$ are nearly indistinguishable on this plot
       beyond $r=1$.  }

      \end{center}
     \end{figure}

In Table~\ref{totalenergy}, we compare our total energies for several closed shell atoms with
the ones from Perdew and Zunger.
In spite of the fact that the minimizing orbitals $\Psi_i$ have a quite
different behaviour from the approximate orbitals calculated by Perdew and
Zunger (which show the behaviour of eigenorbitals in a local potential and look
therefore similar to the orbitals $\Phi_i$ of Fig.1),
the total energies are nevertheless very similar.
In fact the small differences in the total energy in the columns labelled SIC-LDA and PZ
are not due to their approximate solution method but are instead probably due
to the use of an insufficiently dense grid in Ref.~\cite{pz:sic}, since we find the
same differences in comparing the LDA results too.

In Table~\ref{totalenergy}, we also give the results obtained
from the Perdew-Zunger (PZ-LDA) and the Perdew-Wang '92 (PW92-LDA) parametrizations,
the Becke-Lee-Yang-Parr GGA (BLYP) and Hartree Fock (HF).
Of the two LDA functionals, PZ-LDA and PW92-LDA, the latter is the more accurate
parametrization of the correlation energy of a homogeneous electron gas and yields
slightly more accurate total energies.  We include the former only to facilitate comparison
with Ref.~\cite{pz:sic}.
As observed in earlier papers  LDA yields total energies that are not
sufficiently negative while SIC-LDA gives too deep total energies.  The absolute value of the
error is significantly smaller in SIC-LDA than in LDA, but not as small as for BLYP.

In Table~\ref{eigenvalues} we compare the highest occupied eigenvalues of closed shell
atoms.  Here the SIC-LDA functional outperforms all the others.

In Figs. 3, 4 and 5 we show the error in the self consistent charge densities
from the different methods.  The SIC-LDA density is somewhat more accurate than the LDA
density except in a region around 0.3 $a_0$ for Neon.

\begin{table}[tbhp]
\caption[]{Total energies of closed shell atoms in eV. Our results using direct minimization
and the PZ parametrization (SIC-LDA) are compared with the results obtained by Perdew and
Zunger (PZ) for the same functional using an approximate solution method.
The next two columns show the LDA results with the Perdew-Zunger (PZ-LDA) and the
Perdew-Wang '92 (PW92-LDA) parametrizations.
The final three columns are the Becke-Lee-Yang-Parr GGA (BLYP), Hartree Fock (HF) and the
exact value.
The exact values are from Ref.~\cite{Davidson}.
We have used a factor of 27.2112 eV/Hartree for converting most of the energies but
for consistency with Ref.~\cite{pz:sic} a factor of 27.21 eV/Hartree for converting the
SIC-LDA energies.
}
\label{totalenergy}
\vskip 2mm
\begin{tabular}{||l||c|c|c|c|c|c|c||} \hline \hline
    &  SIC-LDA  &     PZ   & PZ-LDA   & PW92-LDA &  BLYP    &    HF    &  exact   \\ \hline
 He &  -79.4    &   -79.4  & -77.1    &    -77.1 &    -79.1 &    -77.9 &    -79.0 \\ \hline
 Be &  -399.9   &   -399.8 & -393.1   &   -393.1 &   -398.9 &   -396.6 &   -399.1 \\ \hline
 Ne &   -3517.9 &  -3517.6 & -3489.1  &  -3489.3 &  -3509.2 &  -3497.9 &  -3508.6 \\ \hline
 Ar &  -14378.8 & -14378.3 & -14307.9 & -14311.6 & -14352.4 & -14335.4 & -14354.6 \\ \hline
\end{tabular}
\end{table}

\begin{table}[tbhp]
\caption[]{Highest occupied eigenvalues of closed shell atoms in eV.
Our results using direct minimization
and the PZ parametrization (SIC-LDA) are compared with the results obtained by Perdew and
Zunger (PZ) for the same functional using an approximate solution method.
The next two columns show the LDA results with the Perdew-Zunger (PZ-LDA) and the
Perdew-Wang '92 (PW92-LDA) parametrizations.
The final three columns are the Becke-Lee-Yang-Parr GGA (BLYP), Hartree Fock (HF) and
experiment.
The experimental values are from Ref.~\cite{Moore}.}
\label{eigenvalues}
\vskip 2mm
\begin{tabular}{||l||c|c|c|c|c|c|c||} \hline \hline
    & SIC-LDA  &    PZ  &PZ-LDA &PW92-LDA&  BLYP  &  HF   &  exper \\ \hline
 He &   -25.8  & -25.8  & -15.5 & -15.5  & -15.8  & -25.0 & -24.6  \\ \hline
 Be &    -9.1  &        & -5.6  &  -5.6  & -5.4   & -8.4  &  -9.3  \\ \hline
 Ne &   -22.8  & -22.9  & -13.5 & -13.5  & -13.2  & -23.1 & -21.6  \\ \hline
 Ar &   -15.9  & -15.8  & -10.4 & -10.4  & -10.0  & -16.1 & -15.8  \\ \hline
\end{tabular}
\end{table}

    \begin{figure}             
     \begin{center}
      \setlength{\unitlength}{1cm}
       \begin{picture}( 8.,7.0)           
        \put(-1.,0.){\includegraphics{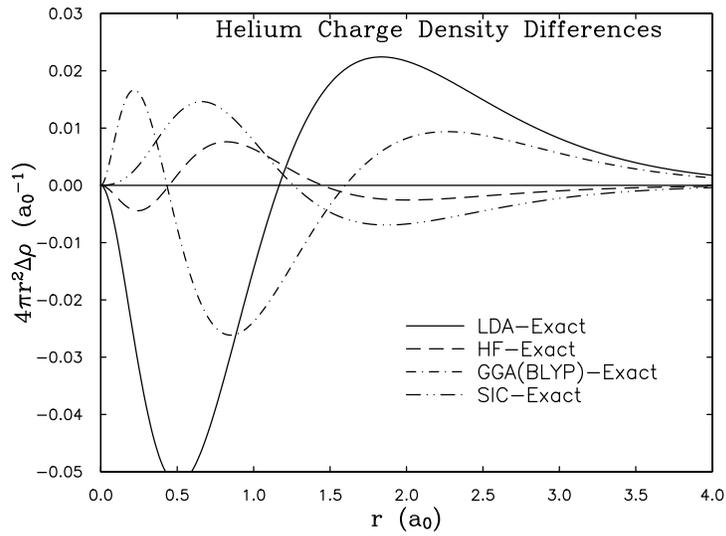} }
       \end{picture}
       \caption{ {Comparison of the charge density of He obtained by different
         methods with the quasi exact charge density from a Hylleras-type
                calculation.} }
      \end{center}
     \end{figure}

    \begin{figure}             
     \begin{center}
      \setlength{\unitlength}{1cm}
       \begin{picture}( 8.,7.0)           
        \put(-1.,0.){\includegraphics{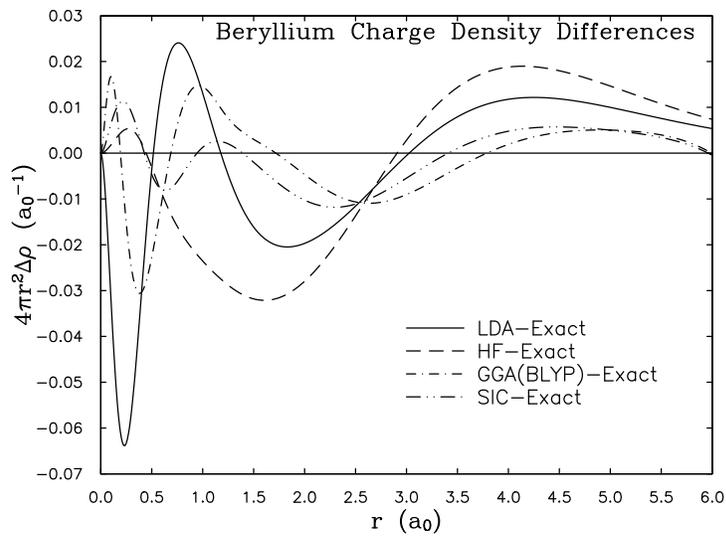}}   
       \end{picture}
       \caption{ {Comparison of the charge density of Be obtained by
                      different methods
                      with the quasi exact charge density from a Quantum Monte Carlo
                calculation.} }
      \end{center}
     \end{figure}

    \begin{figure}             
     \begin{center}
      \setlength{\unitlength}{1cm}
       \begin{picture}( 8.,7.0)           
        \put(-1.,0.){\includegraphics{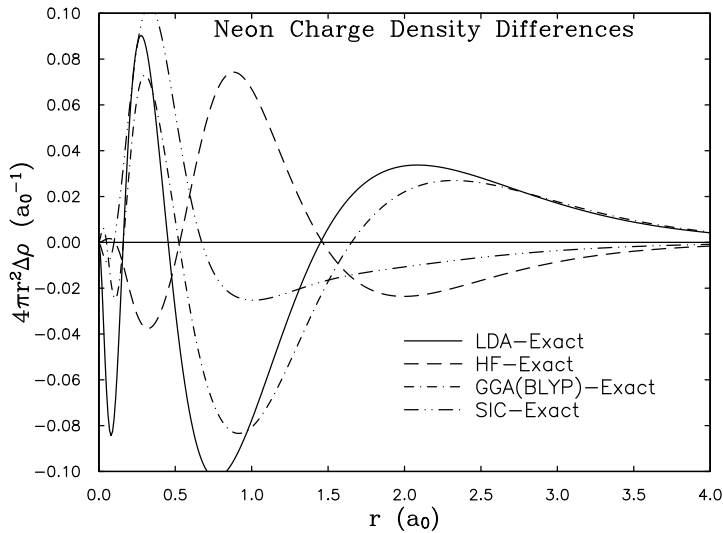}}   
       \end{picture}
       \caption{ {Comparison of the charge density of Ne obtained by
                      different methods
                      with the quasi exact charge density from a Quantum Monte Carlo
                calculation.} }
      \end{center}
     \end{figure}

\section*{Molecular results}
Using pseudo-potentials and plane waves as basis set, we did a series of molecular
calculations. We examined the bond lengths of several small molecules and the
energy released in a chemical reaction. We generated  SIC-LDA
pseudo-potentials according to the procedure described in reference \cite{g:psp}
where however the eigenvalues and charge distributions of  the reference
configuration were the ones from an atomic SIC-LDA calculation. As is to be
expected from the concept of pseudo-potentials representing physical
ions \cite{b:ions} these pseudo-potentials are very similar
to the pure LDA pseudo-potentials
and substituting a LDA pseudo-potential in a SIC-LDA calculation has only a very
minor effect. For consistency, we used however always the SIC-LDA pseudo-potential.
With this kind of pseudo-potential it is
possible to get the correct LDA bond length to within a few thousands of a Bohr
for first row molecules \cite{g:psp}. We expect the same accuracy for
the SIC-LDA pseudo-potentials.
A very attractive feature of the SIC-LDA scheme is that the minimizing orbitals
can usually easily be interpreted in physical terms. They represent either bonds
or lone electron pairs.  In the case of the CH$_4$ molecule for instance, one obtains
4 localized orbitals (each containing a spin up and spin down electron), which are centered
on the 4 lines linking the 4 hydrogens with the central carbon and representing therefore bonds.
In the case of the H$_2$O molecule one obtains again 4 localized orbitals in nearly
tetragonal positions. In this case however only the two sitting on the two lines
linking the O with the two H represent bonds, the other two which are in the half-space
not containing any H are lone electron pairs. In the case of double or triple bonds,
the minimizing functions are banana shaped localized functions distributed around the
line linking the two atoms. We found that in the case of a
Si crystal also, the minimizing orbitals are bond-centered Wannier type functions.
In some cases such as the CO molecule we found two very close minima, the lowest one
corresponding to a triple bond, the other one to a single bond. The charge density
was however very similar in both cases.  Unfortunately, as can be seen from
table~\ref{geometry}, the bond lengths we obtain
for small molecules are systematically too short, and the error was significantly
larger than the error of the LDA bond length.
The error for triple bonds are particularly large.

\begin{table}[tbhp]
\caption[]{Comparison of the LDA and SIC-LDA bond lengths (a.u.) for several
small molecules. The experimental bond lengths (exper.) \cite{dick} and the differences between the
theoretical and experimental bond lengths are given.}
\label{geometry}
\vskip 2mm
\begin{tabular}{||l||c|c|c|c||} \hline
                &  exper.  &SIC error&LDA error& BLYP error\\  \hline\hline
H$_2$           &  1.401   & -.03   &    .05   & .01   \\ \hline
CH$_4$          &  2.052   & -.05   &    .02   & .02   \\ \hline
C$_2$H$_2$ (CH) &  2.005   & -.05   &    .03   &-.01   \\ \hline
C$_2$H$_2$ (CC) &  2.274   & -.09   &   -.01   & .00   \\ \hline
NH$_3$          &  1.912   & -.05   &    .02   & .02   \\ \hline
H$_2$O          &  1.809   & -.05   &    .02   & .03   \\ \hline
BH              &  2.329   & -.09   &    .04   & .01   \\ \hline
LiH             &  3.015   & -.06   &    .01   & .00   \\ \hline
N$_2$           &  2.074   & -.09   &   -.01   & .01   \\ \hline
CO              &  2.132   & -.10   &   -.00   & .02   \\ \hline
CO$_2$          &  2.192   & -.09   &    .00   & .02   \\ \hline
RMS deviation   &          &  .072  &    .024  & .016  \\ \hline
\end{tabular}
\end{table}

Another interesting quantity in this context is the atomization energy,
which is the difference between the molecular total energy and the sum
of the total energies of its constituent atoms. To calculate the energy
of an open-shell atom one has to decide whether one
wants to spherically symmetrize the atom by introducing fractional occupation
numbers.
In the case of GGA is has been found \cite{paint} that the non-spherical atom
gives better results. The theoretical foundation of this empirical observation
is however not quite clear and it can not be predicted which scheme would
give better results in the case of SIC-LDA. To check the accuracy of the SIC-LDA
total energy differences and avoid at the same time the above mentioned symmetry problems
we looked therefore at the energy released in the chemical
reaction 3 H$_2$ + N$_2$ $\rightarrow$ 2 NH$_3$.
All the total energies were calculated after a full relaxation of the atomic positions
within each scheme. In the case of SIC-LDA, we also calculated the energy
difference using the more accurate LDA geometries of the molecules.
As can be seen from Table~\ref{reaction}, the SIC-LDA scheme did worse than the other schemes examined.
By far the best results were obtained with the BLYP \cite{becke} scheme.
To check the accuracy
of the pseudo-potential plane wave method we calculated the LDA result both with
this method and with the Gaussian 94 (G94) \cite{g94} program package,
getting extremely close agreement.
The BLYP calculation was done with the Gaussian 94 software. In all the Gaussian 94 calculations
mentioned in this paper we used a 6-311G++(3df,3pd) basis set.

\begin{table}[tbhp]
\caption[]{The experimental value and theoretical predictions for the
energy (eV) released in the
model chemical reaction 3 H$_2$ + N$_2$ $\rightarrow$ 2 NH$_3$.}
The values are corrected for the zero point energy given in reference \cite{zp}.
\label{reaction}
\vskip 2mm
\begin{tabular}{|c|c|c|c|c|c|} \hline
  exp.   &   LDA  (PSP) &   LDA (G94) &  SIC-LDA  &  SIC-LDA(LDA geom.) & BLYP  \\ \hline
  .76    &  2.1         &  2.1        &  2.6      &  2.9               &  .65  \\ \hline
\end{tabular}
\end{table}

\section*{Conclusions}
The SIC-LDA scheme as proposed by Perdew and Zunger does not give ground state energies
and molecular geometries with sufficient precision. Whereas the atomic results are
superior to the LDA results, the molecular results are clearly worse. A good GGA
scheme such as the BLYP
scheme outperforms both of them in all test cases considered. It is surprising
how well the simple LDA scheme works compared to more sophisticated schemes that
one would expect to give superior results. Charlesworth recently came to similar
conclusions when he systematically examined several weighted density functional
schemes \cite{charles}. All these schemes satisfy sum rules \cite{gun} that are
generally believed to be responsible for the accuracy of the LDA scheme. Thus
it might well be that we actually do not yet fully understand the true reasons
for the success of the LDA scheme.

\section*{Acknowledgments}
We thank  J. Perdew, O. Gunnarson, P. Ballone, M. Parrinello, and A. Zunger for
interesting discussions and useful comments on the manuscript.
CJU is supported by the Office of Naval Research.
Most of the calculations were done on the SP2 supercomputer at the Cornell
Theory center.

\end{document}